\documentclass[aps, twocolumn, groupedaddress,superscriptaddress, showpacs]{revtex4-1}
\usepackage{graphicx}
\usepackage{array}
\usepackage{setspace,transparent}
\usepackage{color}
\usepackage{fontenc,subcaption,ragged2e,amsmath}
\usepackage{mathtools,amssymb,amssymb,nicefrac,tikz}
\usepackage{soul}
\usepackage[font=small,labelfont=bf]{caption}
\usepackage[utf8]{inputenc}
\ifdefined\DeclareUnicodeCharacter
  \DeclareUnicodeCharacter{A3AC}{\nobreakspace}
  \DeclareUnicodeCharacter{2265}{$\geq$}
  \DeclareUnicodeCharacter{21D2}{$\Rightarrow$}
\fi

\DeclareCaptionJustification{justified}{\justifying}
\newcommand{\RN}[1]{%
  \textup{\uppercase\expandafter{\romannumeral#1}}%
}
\DeclareMathAlphabet{\mathpzc}{OT1}{pzc}{m}{it}
\DeclareCaptionJustification{justified}{\justifying}
\begin{document}
\title{Improving robustness of spatial networks via reinforced nodes}
\author{Nir Vaturi}
\affiliation{Department of Physics, Bar Ilan University, Ramat Gan, Israel}
\author{Bnaya Gross}
\affiliation{Department of Physics, Bar Ilan University, Ramat Gan, Israel}
\author{Shlomo Havlin}
\affiliation{Department of Physics, Bar Ilan University, Ramat Gan, Israel}
\date{\today}

\begin{abstract}
Many real-world networks are embedded in space, and their resilience in the presence of reinforced nodes has not been studied. Here we model such networks using a spatial network model that have an exponential distribution of link length $r$ having a characteristic length $\zeta$. We find that reinforced nodes can significantly increase the resilience of the networks which varies with strength of spatial embedding. We also study different reinforced node distribution strategies for improving the network resilience. Interestingly, we find that the best strategy is highly dependent on the stage of the percolation process, i.e., the expected fraction of failures. Finally, we show that the reinforced nodes are analogous to an external field in percolation phase transition i.e., having the same critical exponents and that the critical exponents satisfy Widom's relation. 
\end{abstract}

\maketitle
\section{Introduction}
Network theory has made a significant contribution to understanding the complexity of different systems, such as communication networks \cite{lambiotte2008geographical}, transportation systems \cite{zeng2019switch} and even neuronal networks \cite{fornito2016fundamentals, breakspear2017dynamic}. The resilience of such networks is often studied under a percolation process \cite{essam1980percolation, stauffer2018introduction, kirkpatrick1973percolation, bunde2012fractals} where a fraction $1 - p$ of nodes is removed from the network and the size of the largest connected component, $P_{\infty}$ is evaluated. It has been found that these networks experience a second-order phase transition at a critical point $p_c$ \cite{cohen2000resilience}. Below this transition point, the size of the giant component is zero and non-zero above it. The giant component is used to describe the network functionality where nodes that are connected to it are considered functional, while isolated clusters are considered non-functional.

Many real-world networks are spatially embedded \cite{barthelemy2011spatial, watts1998collective, penrose2003random, grosspercolation, gross2017multi}, and it has been shown that the spatiality of a network affects the phase transition of the percolation process \cite{danziger2016effect}. Examples of such spatial networks are infrastructure networks \cite{hines2010topological}, brain network \cite{barthelemy2011spatial} and transportation networks \cite{strano2017scaling, weiss2018global, li2015percolation}. Percolation processes have been studied thoroughly for complex networks \cite{dorogovtsev2008critical, cohen2000resilience, cohen2021percolation, callaway2000network}, but the robustness have not been studied for spatial networks in the presence of reinforced nodes.

Recently, the question of centralization vs. decentralization of infrastructures has been emerged, which led to the idea of reinforced nodes. These reinforced nodes can function and support other nodes in their cluster even if they are not connected to the giant component \cite{yuan2017eradicating, kfir2022optimization} and can be significant for the robustness of real-life networks. For example, in the case of the internet, satellites \cite{henderson1999transport} can also be used in order to exchange information, meaning they are functional without being directly connected to the giant component. Another example are power-grids, where reinforced nodes represent generators, each having their own source of energy and being able to support themselves and their clusters.

In this paper, we will study the effects of reinforced nodes on spatial networks and how to optimally distribute them.

\maketitle
\section{model}
Our model consists of a $2D$ lattice of size $ N = L \times L$ initially without links. Next, we generate the length of the links, $r$, between the nodes which follow an exponential distribution \cite{gross2017multi,danziger2016effect,bonamassa2019critical,gross2021interdependent,gotesdyner2022percolation},
\begin{equation}
    P(r) \sim e^{-r/ \zeta} \quad.
    \label{eq:1}
\end{equation}
The process of creating a spatial network requires 3 successive steps. The first step is to randomly choose a node. The second step is to randomly draw a link length from the distribution in Eq. (1). The third step is to randomly choose an angle and identify the closest target node for the link and create an edge. For our specific model, we chose the average degree (number of links) of the nodes to be $\langle k \rangle = 4$. In order to achieve a given $\langle k \rangle$ we repeat these steps $N \cdot \langle k \rangle / 2$ times.

\begin{figure}[h!]
	\centering
	\begin{tikzpicture}[      
	every node/.style={anchor=north east,inner sep=0pt},
	x=1mm, y=1mm,
	]   
	\node (fig1) at (-30,0)
	{\includegraphics[scale=0.345]{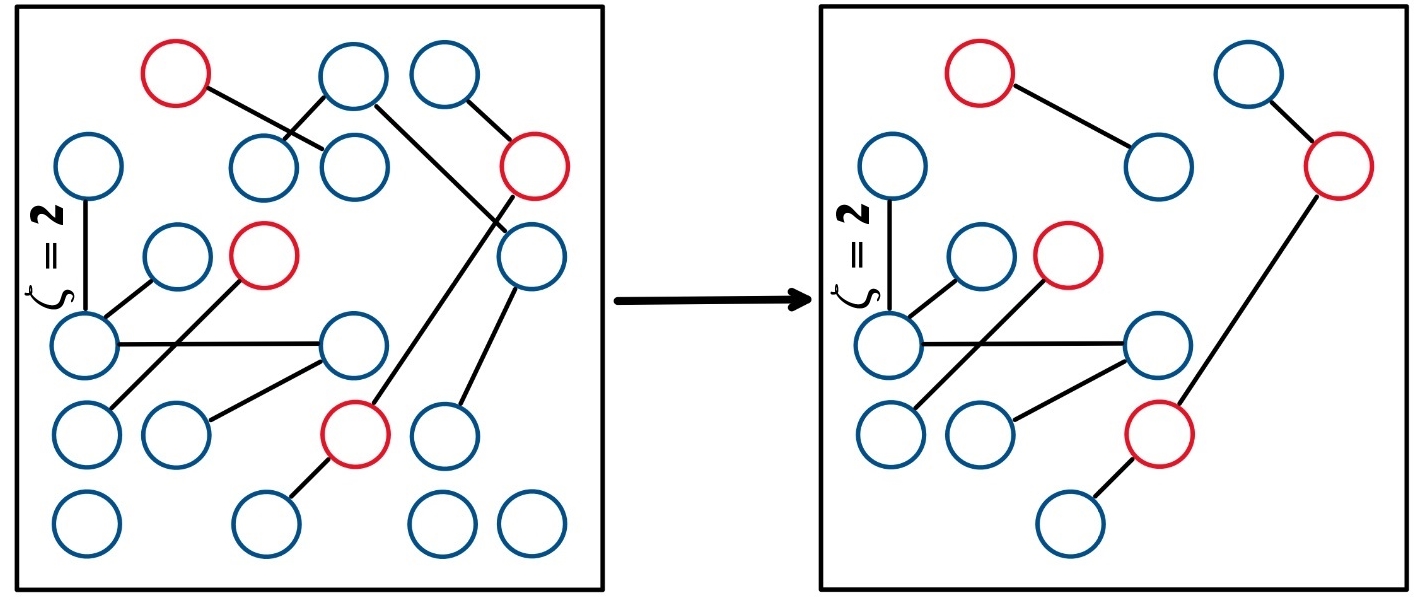}};
	\end{tikzpicture}
	\caption{\textbf{Illustration.} The nodes are placed as the sites of a $2D$ lattice, while the links are added according to Eq.~\eqref{eq:1}. The characteristic length of the links is $\zeta$ (left link in both boxes) while a small fraction $\rho$ of the nodes are randomly reinforced (red nodes). (left box) The network at a certain point of the percolation process, contains the giant component and finite clusters. (right box) $P_{\infty}$ represents the functioning nodes in the network, i.e., it includes all nodes that are part of the giant component or connected to a finite cluster with at least one reinforced node, total 13 nodes. }
	\label{fig:Illustration}	
\end{figure}

It is important to note, that for values of $\zeta \to 0$ our model generates a $2D$ lattice, where each node is connected only to its nearest neighbors, and has a known value of $p_c \simeq 0.59$ \cite{grosspercolation}. While for values of $\zeta \to \infty$ our model generates an $ER$ network, where all links have a pre-determent probability of being cast, and has a known value of $p_c = 1 / \langle k \rangle = 1 / 4 = 0.25$ \cite{grosspercolation}.

After generating the spatial network, we chose a fraction $\rho$ of the nodes of the network to be reinforced nodes (see Fig.~1). The reinforced nodes are chosen randomly, with the only condition that the nodes are part of the original giant component of the network, i.e., at $p = 1$. We now use the notation $P_{\infty}$ as the fraction of \textit{functioning} nodes in the network and we analyze $P_{\infty}$ as a function of $p$, where $1 - p$ is the fraction of non-removed nodes from the network.

\maketitle
\section{Results}

The resilience of spatial networks can be studied using percolation process, as shown in Fig.~2 for the model above, for different values of $\zeta$ and different values of $\rho$. We included only 2 values of $\zeta$ in the figure, other values of $\zeta$ show similar results and can be seen in the appendix. As we can see, the existence of reinforced nodes makes the spatial network more resilient to random failures, and the phase transition is removed. It can also be seen that the higher the fraction of reinforced nodes, the more resilient is the network. And lastly, the shorter is $\zeta$, i.e., the stronger is spatiality, the higher is the impact of the reinforced nodes on the network.

\begin{figure}[h!]
	\centering
	\begin{tikzpicture}[      
	every node/.style={anchor=north east,inner sep=0pt},
	x=1mm, y=1mm,
	]   
	\node (fig1) at (-30,0)
	{\includegraphics[scale=0.215]{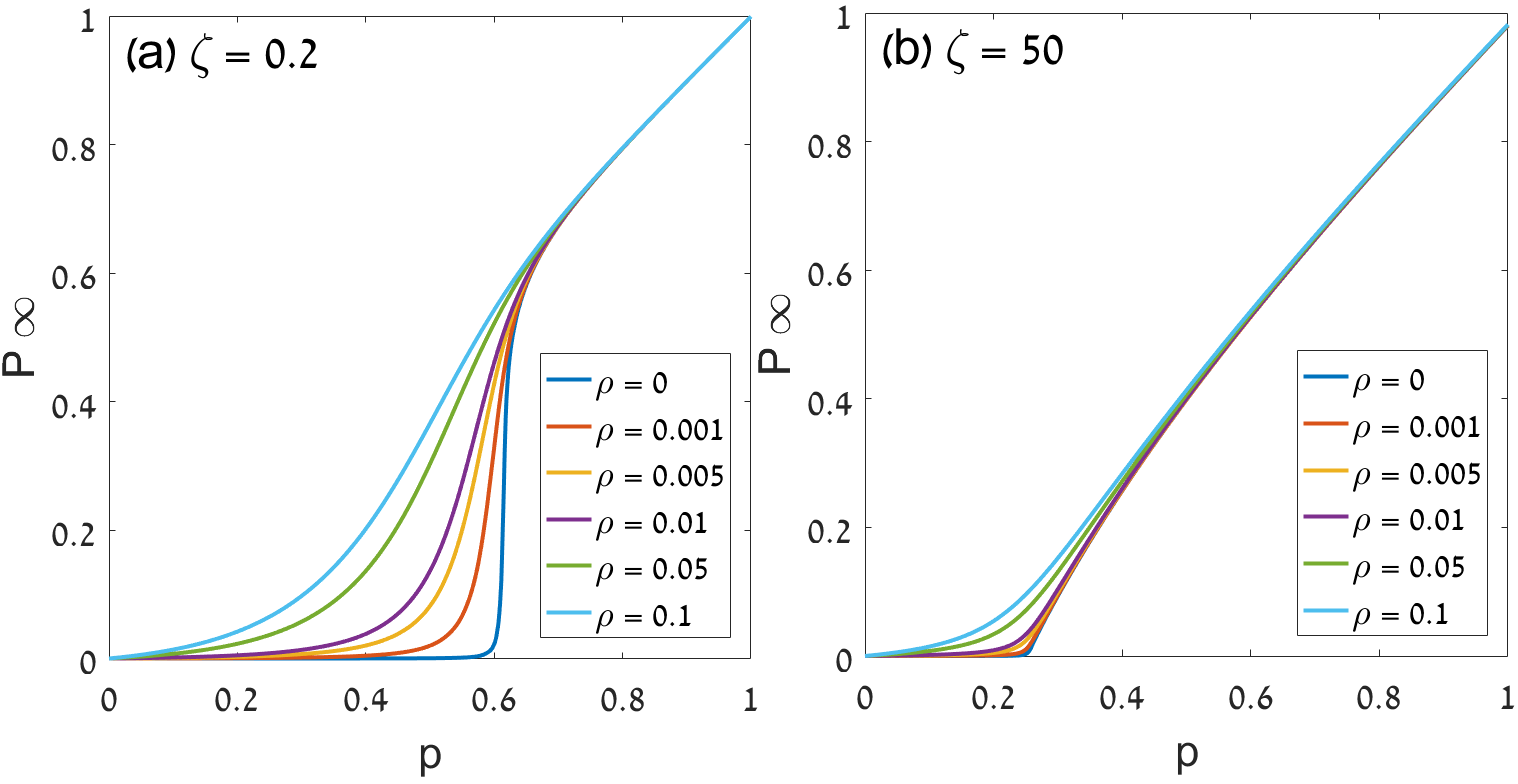}};
	\end{tikzpicture}
	\caption{\textbf{Randomly distributed reinforced nodes in spatial networks.} \textbf{(a)} The giant component, $P_{\infty}$, as a function of $p$ for $\zeta = 0.2$ (similar to $2D$ lattice). The phase transition at $p_c \simeq 0.59$ is being removed by even a small fraction of reinforced nodes. The network becomes more resilient as the fraction of reinforced nodes increases. \textbf{(b)} The giant component, $P_{\infty}$, as a function of $p$ for $\zeta = 50$. As $\zeta$ increases, we can see similar results but a significantly weaker effect compared to (a). Note, due to large $\zeta$, the phase transition is close to the $ER$ limit of $p_c = 0.25$ (here $\langle k \rangle = 4$). }
	\label{fig:r1}	
\end{figure}

Once we establish that reinforced nodes increase significantly the resilience of spatial networks, we can address the question of identifying better strategies, i.e., we can ask what is the best strategy to distribute the reinforced nodes to maximize the network resilience? To do so, we repeated the percolation process for different values of $\zeta$ and different values $\rho$ testing 6 different distribution strategies:
\begin{itemize}
  \item Random distribution, as shown in Fig.~2.
  \item Nodes with the highest degree $k$.
  \item Nodes with the lowest degree $k$.
  \item Nodes with the longest average link length.
  \item Nodes with the shortest average link length.
  \item Node with highest weighted degree, defined as:
  \end{itemize}
\begin{equation}
    w_i = \sum_{j = 1}^{N} A_{ij} \sqrt{(x_i - x_j)^2 + (y_i - y_j)^2}
    \label{wi_eq}
\end{equation}
where $A_{ij}$ is the adjacency matrix.
The weighted degree strategy is a combination of the four previous strategies since it takes into consideration both the average length of the links and the degree $k$ of the node.

\begin{figure}[h!]
	\centering
	\begin{tikzpicture}[      
	every node/.style={anchor=north east,inner sep=0pt},
	x=1mm, y=1mm,
	]   
	\node (fig1) at (-30,0)
	{\includegraphics[scale=0.13]{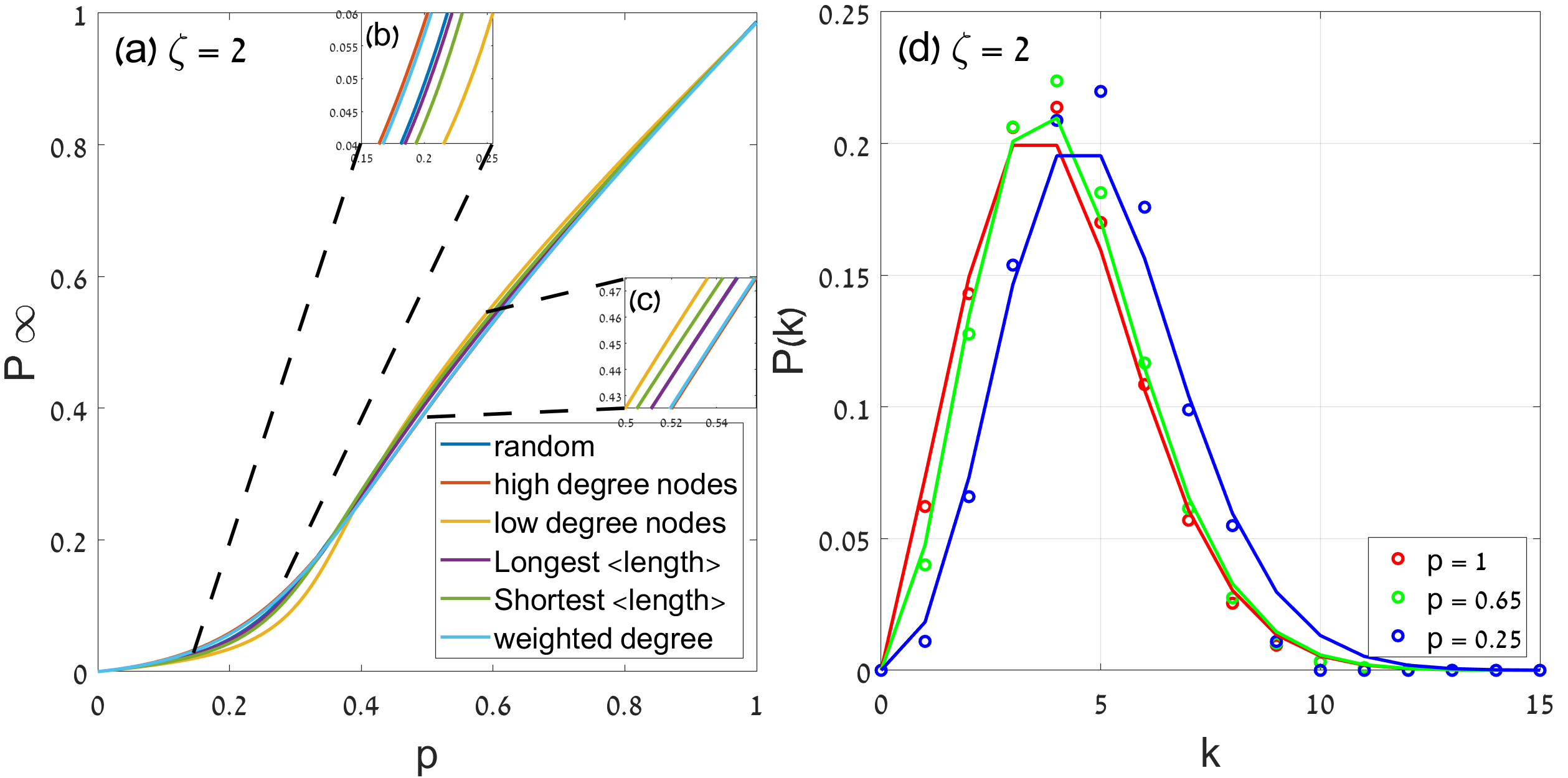}};
	\end{tikzpicture}
	\caption{\textbf{Prioritize strategies for reinforced nodes distribution.} \textbf{(a)} We tested $P_\infty$ as a function of $p$ for 6 different strategies of reinforced nodes distribution: random distribution, high and low degree node preference, longest and shortest average length preference, and weighted degree preference. Here $\zeta = 2$. \textbf{(b)} Zoom in of the lower values $p$ (below $p_c$). We find that for all values of $\zeta$ (like here for $\zeta = 2$) and for low values of $p$ (close to the percolation threshold), the best nodes to reinforce are the nodes with the highest degree, and the worst nodes to reinforce are the nodes with the lowest degree. \textbf{(c)} Zoom in for the high values $p$ (above $p_c$). For high values of $p$ the statement in (b) is reversed, but the difference becomes very small and almost does not impact the network robustness for the same $\zeta$ value compared to the case of no reinforced nodes. Here $\rho = 0.1$. \textbf{(d)} Shows the degree distribution of the nodes having degree $k$ in the giant component. Here $\zeta = 2$ and $\rho = 0$. As can be seen, the distribution shifts to the right for low values of $p$, i.e., towards larger degree values. Thus, at early stages of percolation, it is more useful to reinforce low degree nodes since they fail more frequently, while at small values of $p$ it is better to reinforce the high degree nodes. We can see that our simulation results (circles) and the analytic MF solution (the continuous lines), are in good agreement, indicating that although the analytical solution is based on MF theory, it catches also quite well the behaviour of spatial networks. }
	\label{fig:bestrho}	
\end{figure}

As we can see in Fig.~3(a) - (c), the functionality of the network changes depending on the distribution strategy of the reinforced nodes. For high values of $p$, the best distribution strategy of reinforced nodes is low degree nodes. For low values of $p$, the best distribution strategy is to reinforce high degree nodes. These results are shown for $\zeta = 2$, but valid also for other values of $\zeta$ (see appendix). In order to understand why the percolation process has different best distribution strategies for reinforced nodes at different stages, we study the distribution of the degree $k$ in the giant component for different stages in the percolation (different values of $p$). We find, as shown in Fig.~3(d), that the distribution is shifting for lower values of $p$ towards large degrees, which means that at the earlier stages of the percolation, low degree nodes have a much higher probability to be disconnected from the giant component. Thus, at early stages of percolation, making low degree nodes that have high probability of failing as reinforced nodes improve the robustness, while at later stages, high degree nodes are preferred to be chosen as reinforced nodes.

\maketitle
\section{test on a real network: the EU power grid}
To further validate our results, we studied the impact of reinforced nodes on the EU power grid network. The edge distribution of this network was found to follow Eq.~\eqref{eq:1} \cite{danziger2016effect}. The number of nodes in the network is $N = 1254$, while the average degree is $\langle k \rangle = 1.44$ \cite{danziger2016effect}. The size here is much smaller compared to our simulations before, thus, we expect the results to be noisy and, due to the lower degree, higher value of $p_c$.

\begin{figure}[h!]
	\centering
	\begin{tikzpicture}[      
	every node/.style={anchor=north east,inner sep=0pt},
	x=1mm, y=1mm,
	]   
	\node (fig1) at (-30,0)
	{\includegraphics[scale=0.2]{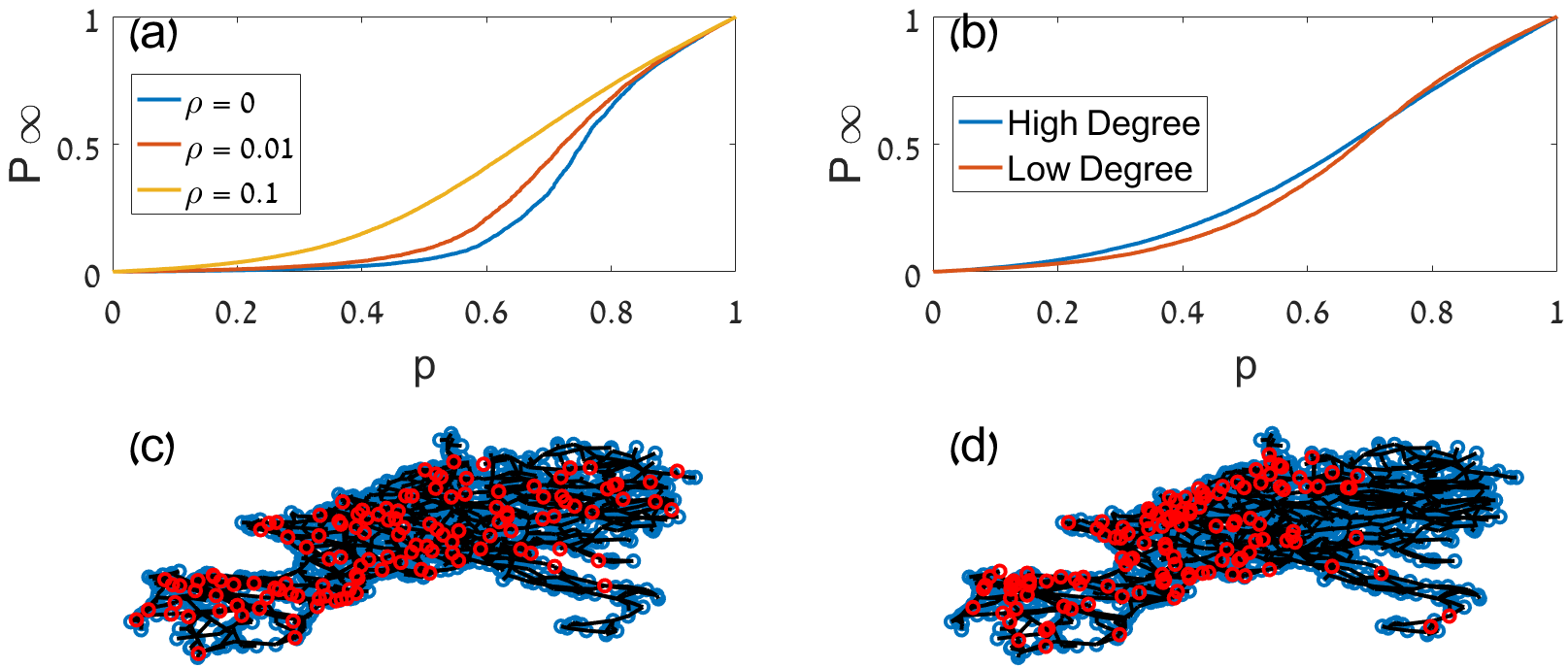}};
	\end{tikzpicture}
	\caption{\textbf{The power grid network of Europe.} \textbf{(a)} Randomly distributed reinforced nodes in the power grid network of Europe \cite{danziger2016effect}. As we can see, the small amount of nodes $N$ in the network makes the percolation transition broad, but we can see that $p_c \simeq 0.7$ (which is reasonable due to the low value of the average degree of the network $\langle k \rangle$ as we show in the appendix). We can also see that the higher the fraction of reinforced nodes, the more the network becomes more resilient, similar to our model. \textbf{(b)} The effects of different reinforced node strategies in the power grid network of Europe on the percolation of the network. We can see the percolation process of the power grid network of Europe with low degree nodes and high degree nodes strategies. Similar to Fig 3, we can clearly see that for the power grid of Europe network we find the low degree node strategy improve the robustness at earlier stages of the percolation, while for later stages of the percolation, the high degree node strategy is clearly better. \textbf{(c)} An illustration of the EU power grid network with reinforced nodes on the nodes with the highest degrees $k$. \textbf{(d)} An illustration of the EU power grid network with reinforced nodes on the nodes with the lowest degrees $k$. For both (c) and (d), the blue circles represent the regular nodes, the red circles represent the reinforced nodes, and the black lines are the links between the nodes in the network. For (b) - (d) $\rho = 0.1$}
	\label{fig:EU}	
\end{figure}

In Fig.~4(a) we demonstrate the impact of randomly distributed reinforced nodes in a real world network, the EU power grid, for different values of $\rho$. As one can see, we get similar results as those found in our model. As shown in Fig.~2, the presence of a small fraction of reinforced nodes makes the network significantly more resilient. In Fig.~4(b) we show the percolation process of the power grid of Europe network with 2 strategies according to Fig.~3, low degree distribution and high degree distribution. We can clearly see similar results to our model. For early stages of the percolation process, it is better to reinforce the nodes with the lower degree, while for later stages of the percolation, it is better to reinforce the nodes with the higher degree. In Fig.~4(c) - (d) we can see the power grid network of Europe with low degree distribution and high degree distribution strategies.

\maketitle
\section{External field analogy}

Here we argue that the concentration $\rho$ of reinforced nodes is analogous to external field in percolation \cite{stauffer2018introduction, bunde2012fractals}. The key critical exponents $\beta$, $\delta$, and $\gamma$ that describe the behavior of the system near criticality will be derived below \cite{bonamassa2019critical, fan2018structural}. These key critical exponents also fulfill Widom's identity $\delta - 1 = \gamma / \beta$ \cite{stauffer2018introduction, bunde2012fractals, gross2020interconnections, yuan2017eradicating, fan2018structural}.

The critical exponent $\beta$ describes the behavior of the order parameter $P_{\infty}$ near the critical point with zero-field ($\rho = 0$, i.e., no reinforced nodes) and is given by
\begin{equation}
    P_{\infty}(0,p) \sim \left( p - p_c \right) ^{\beta} \quad.
    \label{beta_eq}
\end{equation}
\textit{At the critical point,} ($p = p_c$), the increase of the order parameter with the magnitude of the field, i.e., the concentration of reinforced nodes $\rho$, is expected to yield the critical exponent $\delta$ as
\begin{equation}
    P_{\infty}(\rho,p_c) \sim \rho^{1/\delta} \quad.
\end{equation}
The susceptibility of the system, $\chi$, is given by the partial derivative of the order parameter with respect to the field, $\rho$, and scales near the critical point with the exponent $\gamma$ as 
\begin{equation}
    \chi \equiv \left(\frac{\partial P_{\infty}(\rho,p)}{\partial \rho}\right)_{\rho \rightarrow 0} \sim |p - p_c|^{-\gamma} \quad.
    \label{gamma_eq}
\end{equation}

\begin{figure}[h!]
	\centering
	\begin{tikzpicture}[      
	every node/.style={anchor=north east,inner sep=0pt},
	x=1mm, y=1mm,
	]   
	\node (fig1) at (-30,0)
	{\includegraphics[scale=0.105]{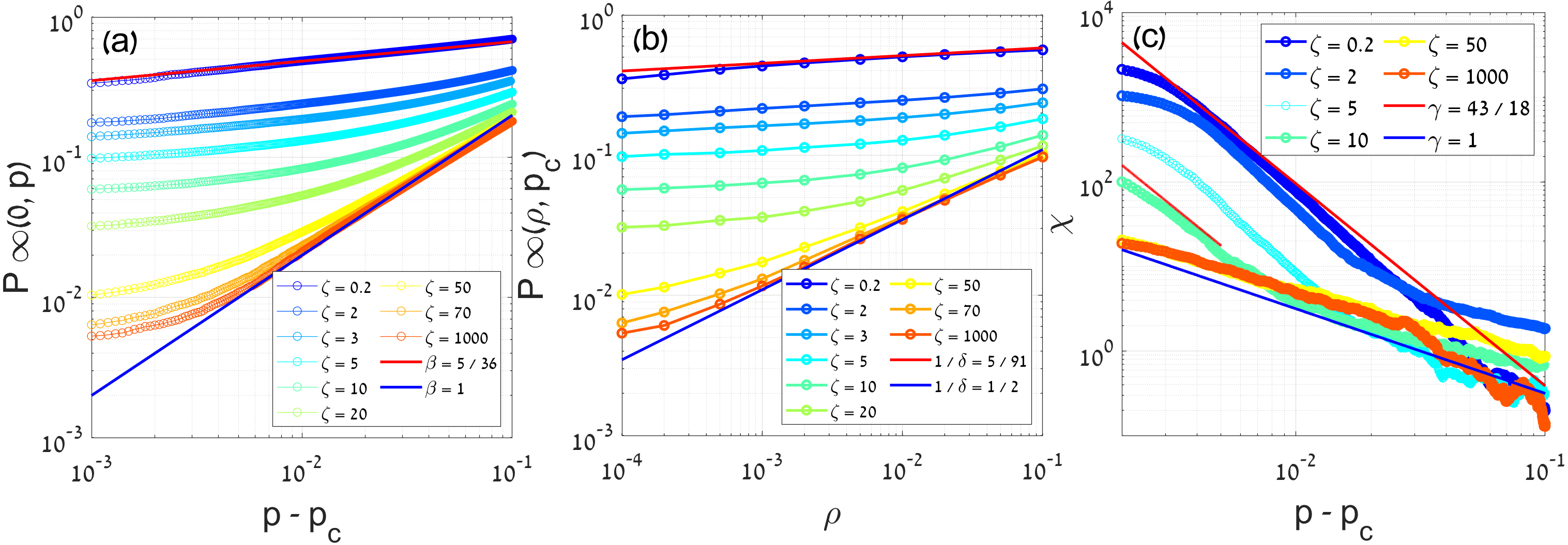}};
	\end{tikzpicture}
	\caption{\textbf{Critical exponents and external field analogy.} \textbf{(a)} The critical exponent $\beta$. One can see a crossover between the known value of $\beta$ for $2D$ lattice ($\zeta = 0.2$), $\beta = 5 / 36$, and the known value of $\beta$ for $ER$ network ($\zeta = 1000$), $\beta = 1$. For intermediate values of $\zeta$, one can see clearly a crossover between these two exponents on the same curve. Close to $p_c$ (left) where the correlation length is large (above $\zeta$), one can see  the $2D$ lattice $\beta$ exponent, while far from $p_c$ (right) where the correlation length is short (below $\zeta$), one can see the $ER$ $\beta$ exponent. \textbf{(b)} The critical exponent $\delta$. Similar to $\beta$, there is a crossover between the known value of $\delta$ for $2D$ lattice ($\zeta = 0.2$), $\delta = 91 / 5 $, and the known value of $\delta$ for $ER$ network ($\zeta = 1000$), $\delta = 2$. \textbf{(c)} The critical exponent $\gamma$. Also here there is a crossover between the known value of $\gamma$ for $2D$ lattice ($\zeta = 0.2$), $\gamma = 43 / 18$, and the known value of $\gamma$ for $ER$ network ($\zeta = 1000$), $\gamma = 1$.}
	\label{fig:CE3}	
\end{figure}

As we can see in Fig.~5, all critical exponents show a crossover between their known values for the $2D$ lattice and the $ER$ network as $\zeta$ changes \cite{bonamassa2019critical}. For strong spatiality (i.e. $\zeta = 0.2$), we obtain $\beta = 5 / 36$, $\delta = 91 / 5$, and $\gamma = 43 / 18$, while for weak spatiality (i.e. $\zeta = 1000$), we find $\beta = 1$, $\delta = 2$, and $\gamma = 1$, which are the known values for these critical exponents for both of these networks. Both sets of critical exponents satisfy Widom's identity $\delta - 1 = \gamma / \beta$ \cite{stauffer2018introduction, bunde2012fractals}.

\section{Summery}
We showed that reinforced nodes have a significant effect on the resilience of spatial networks. Even a small fraction of reinforced nodes destroy the phase transition of the percolation process and increase the functional component. We showed that the higher the fraction of reinforced nodes, the more resilient is the network, i.e., increase the functional component. We also found that the lower the value of $\zeta$, the stronger is the effect for the same fraction of reinforced nodes on the resilience of the network. We showed that the best reinforced node distribution strategy highly depends on the stage of the percolation process. For earlier stages of the percolation process, low $1-p$, it is better to reinforce the low degree nodes, while for later stages of the percolation process, it is better to reinforce the high degree nodes. We also showed the effect of reinforced nodes in a real-world spatial embedded network. Finally, we showed that reinforced nodes are analogous to external field also in the presence of spatial constraints with a crossover phenomenon for intermediate values of $\zeta$ and that the critical exponents that we found satisfy the Widom's identity $\delta - 1 = \gamma / \beta$.

\section{Acknowledgments}
We thank the Israel Science Foundation, the Binational Israel-China Science Foundation Grant No.\ 3132/19, ONR, NSF-BSF Grant No.\ 2019740, the EU H2020 project RISE (Project No. 821115), the EU H2020 DIT4TRAM, and DTRA Grant No.\ HDTRA-1-19-1-0016 for financial support.

\appendix
\counterwithin{figure}{section}
\section{The effect of reinforced nodes for different values of $\zeta$}
In this appendix section, we will show random distributed reinforced nodes for other values of $\zeta$(i.e $\zeta = 10$ and $\zeta = 1000$). As can be seen in Fig.~A1, we get similar results as in Fig.~2 in the main text, and the existence of reinforced nodes makes the spatial network more resilient to random failures, and the phase transition is removed. The meaning is that those results apply for all values of $\zeta$ since we cover the entire range from $\zeta = 0.2$ (Fig.~2(a) in the main text) to $\zeta = 1000$ (Fig.~A1(b)). 
\begin{figure}[h!]
	\centering
	\begin{tikzpicture}[      
	every node/.style={anchor=north east,inner sep=0pt},
	x=1mm, y=1mm,
	]   
	\node (fig1) at (-30,0)
	{\includegraphics[scale=0.21]{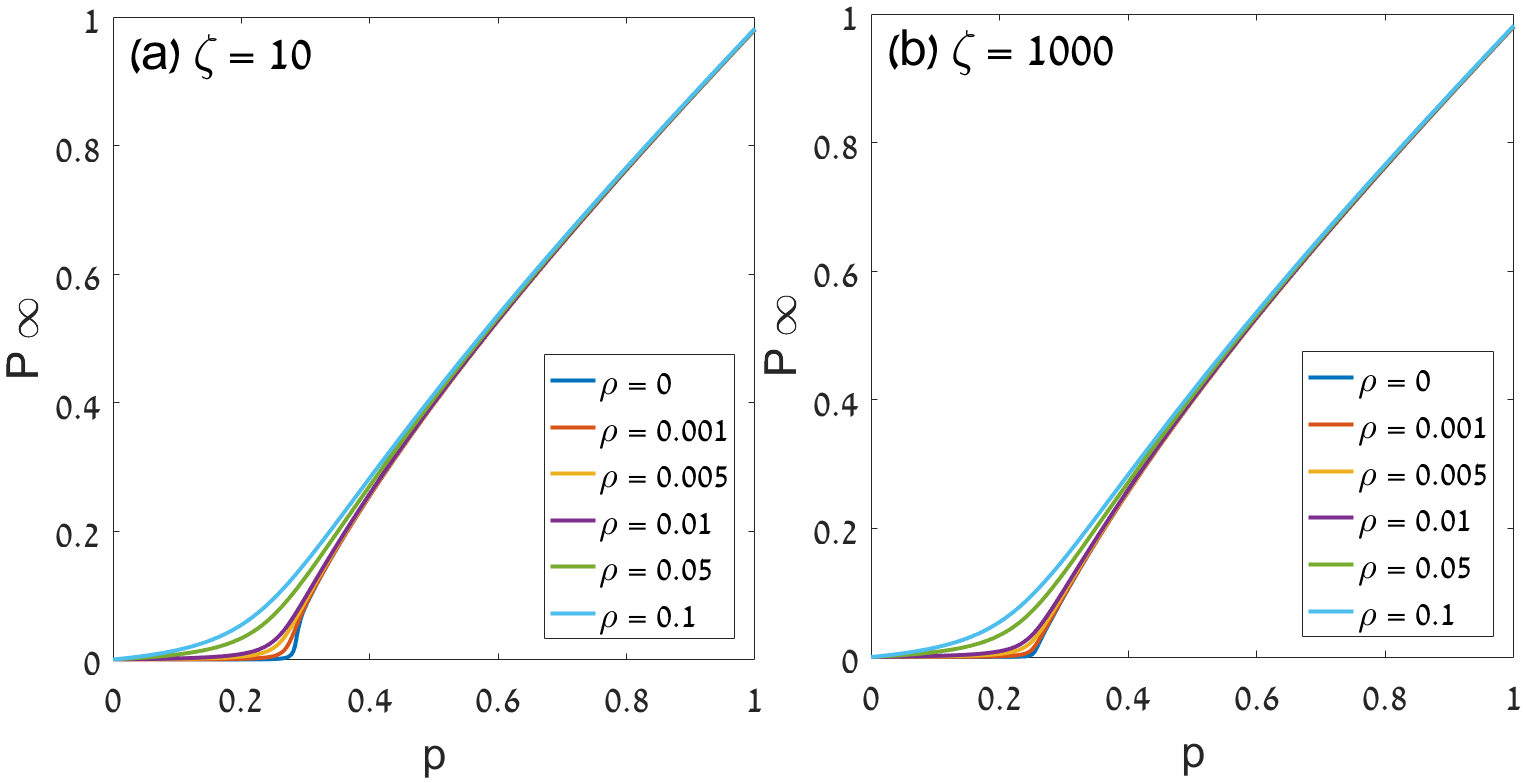}};
	\end{tikzpicture}
	\caption{\textbf{Randomly distributed reinforced nodes in spatial networks.} \textbf{(a)} The giant component, $P_{\infty}$, as a function of $p$ for $\zeta = 10$. \textbf{(b)} The giant component, $P_{\infty}$, as a function of $p$ for $\zeta = 1000$. We can see that for all values of $\zeta$ analysed we get qualitatively similar results to those shown in Fig.~2 in the main text. }
	\label{fig:a1}
\end{figure}

\section{Prioritized distribution of reinforced nodes for different values of $\zeta$}
In this appendix section, we will identify preferred strategies for reinforced distribution for other values of $\zeta$. As we can see in Fig.~B1, for other values of $\zeta$ we obtain similar results to those seen in Fig.~3 in the main text. We see also here that different reinforced node distributions are preferred for different stages of the percolation process. While the low degree nodes are better to reinforce at the earlier stages of the percolation process (large $p$), it is better to reinforce the high degree nodes at later stages of the percolation process (low $p$). These similarities suggest that our results are valid for the entire $\zeta$ range. Notice that the lowest value of $\zeta$ that we included here is $\zeta = 2$ instead of $\zeta = 0.2$ in Fig. 2. This is since for $\zeta = 0.2$ the degree and edge distributions are similar to those of $2D$ lattice i.e. $\delta_{k,4}$ for the degree distribution and all four nearest neighbors are edges. In this case all methods will show similar behaviour.

\begin{figure}[h!]
	\centering
	\begin{tikzpicture}[      
	every node/.style={anchor=north east,inner sep=0pt},
	x=1mm, y=1mm,
	]   
	\node (fig1) at (-30,0)
	{\includegraphics[scale=0.175]{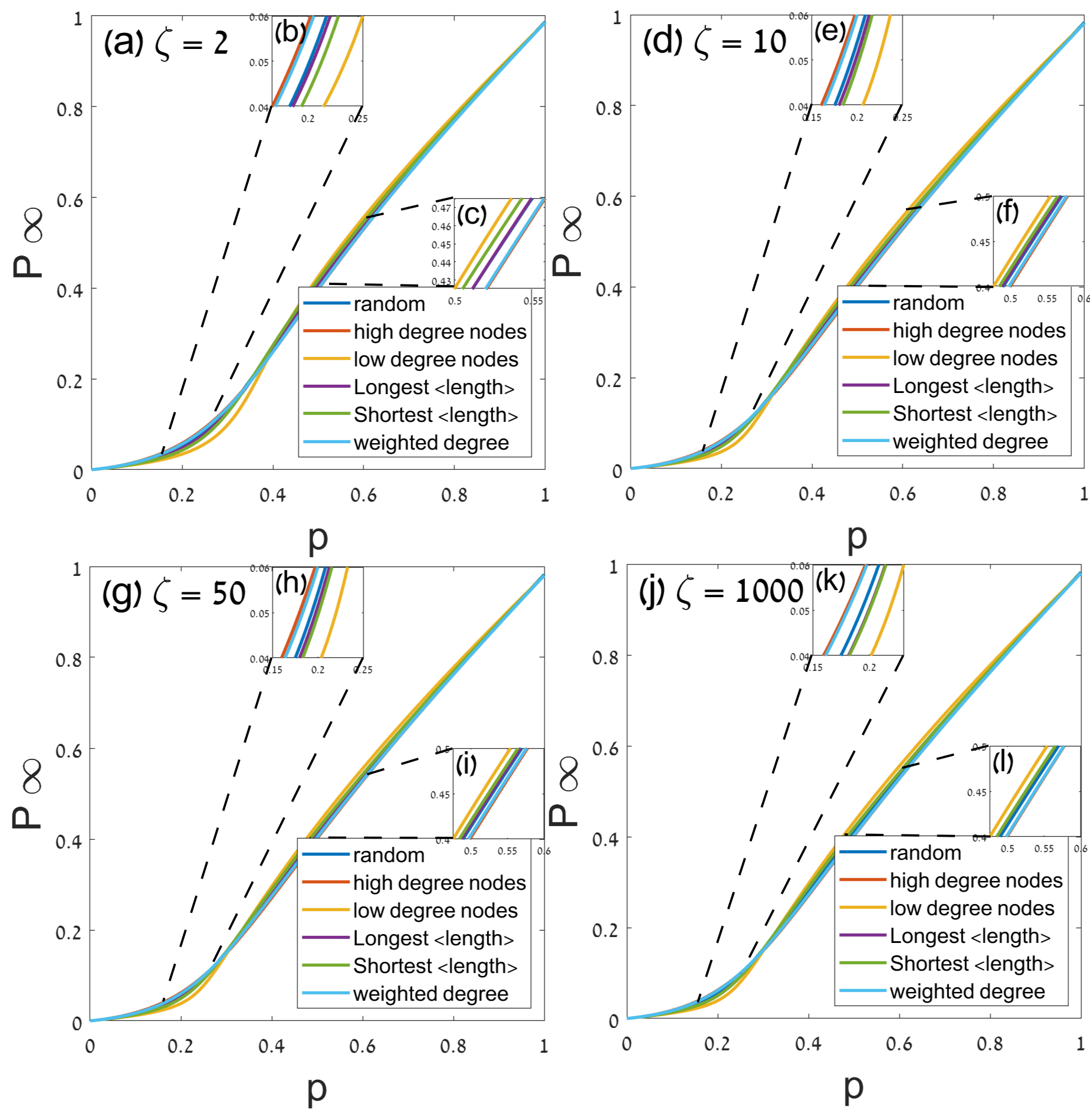}};
	\end{tikzpicture}
	\caption{\textbf{Prioritize strategies for reinforced distribution.} \textbf{(a) - (c)} The giant component, $P_{\infty}$, as a function of $p$ for $\zeta = 2$. \textbf{(d) - (f)} The giant component, $P_{\infty}$, as a function of $p$ for $\zeta = 10$. \textbf{(g) - (i)} The giant component, $P_{\infty}$, as a function of $p$ for $\zeta = 50$. \textbf{(j) - (l)} The giant component, $P_{\infty}$, as a function of $p$ for $\zeta = 1000$. Here $\rho = 0.1$. We can see that for all values of $\zeta$, we get similar results as those shown in Fig.~3(a) - (c) in the main text.}
	\label{fig:bestrho4ever}	
\end{figure}

\section{Analytical results for the degree distribution of the nodes in the giant component during percolation process}
In this section, we will solve analytically the node degree distribution in the giant component for $ER$ networks.

We start by defining two events: $A$ is the event of a node being of degree $k$ and $B$ is the event of a node being in the giant component after the random removal of a fraction of $1-p$ of nodes. Thus, the degree distribution of nodes in the giant component will be obtained by the conditional probability $P(A|B)$ which according to the Bayes' theorem is
\begin{equation}
    P(A|B) = \frac{P(B|A) P(A)}{P(B)}.
\end{equation}

The probability for a node to have degree $k$ is
\begin{equation}
    P(A) = p_k.
\end{equation}

The probability of a node being a part of the giant component is
\begin{equation}
    P(B) = P_{\infty}.
\end{equation}

The conditional probability is:
\begin{equation}
    P(B|A) = p(1 - u^k)
\end{equation}

where $u$ is the probability that a node at the end of an edge of a randomly chosen node is \textit{not} in the giant component \cite{newman-pre2001}. Thus,

\begin{equation}
    P(A|B) = \frac{p(1 - u^k) p_k}{P_{\infty}}.
\end{equation}

In the case of ER, $P_{\infty} = 1-u$ \cite{newman-pre2001} and $p_k = \frac{\langle k \rangle^k e^{-\langle k \rangle}}{k!}$. Thus, 

\begin{equation}
    P(A|B) = \frac{p(1 - (1 - P_{\infty})^k) \langle k \rangle^k e^{-\langle k \rangle}}{P_{\infty} k!}
    \label{eq:2}
\end{equation}

where \cite{er1959},
\begin{equation}
    P_{\infty} = p(1 - e^{-\langle k \rangle P_{\infty}}).
    \label{eq:3}
\end{equation}

In Fig.~C1 we can see the degree distribution of nodes in the giant component at different stages of the percolation process and for different values of $\zeta$. Here we can see the simulation results (circles) and the analytic solution for ER (continuous lines) from Eq.~\eqref{eq:2}. As seen already in Fig.~3 in the main text, we see here also similar behaviour and good agreement between the simulation results and the analytic solution (of mean-field ER) for all values of $\zeta$ indicating that although the analytical solution is mean-field it catches the behaviour of spatial networks.

\begin{figure}[h!]
	\centering
	\begin{tikzpicture}[      
	every node/.style={anchor=north east,inner sep=0pt},
	x=1mm, y=1mm,
	]   
	\node (fig1) at (-30,0)
	{\includegraphics[scale=0.17]{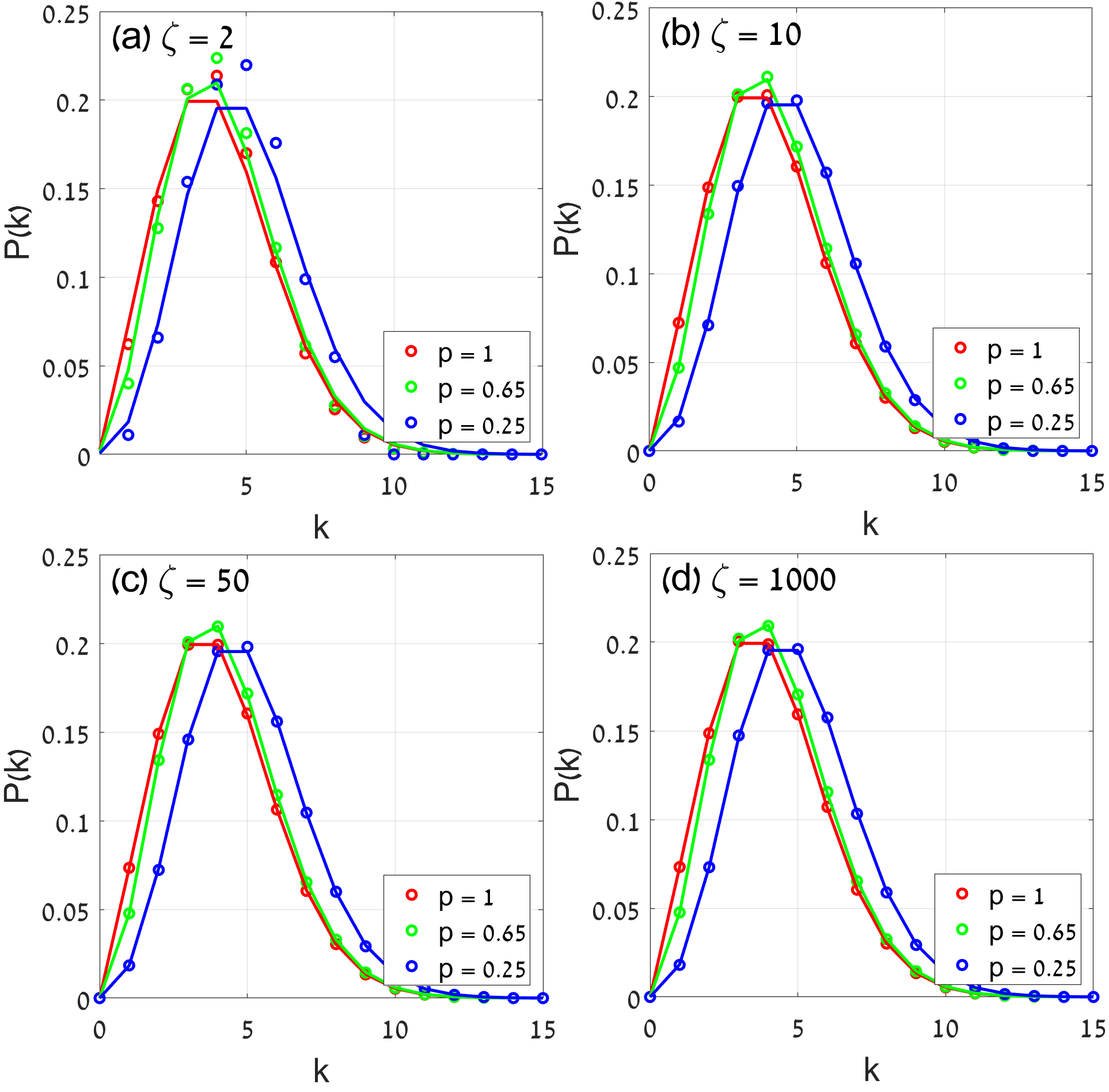}};
	\end{tikzpicture}
	\caption{\textbf{Degree distribution of nodes in the giant component (for $\mathbf{\langle k \rangle = 4}$).} \textbf{(a)} The degree distribution of nodes, $P(k)$, as a function of $k$ for $\zeta = 2$. \textbf{(b)} The degree distribution of nodes, $P(k)$, as a function of $k$ for $\zeta = 10$. \textbf{(c)} The degree distribution of nodes, $P(k)$, as a function of $k$ for $\zeta = 50$. \textbf{(d)} The degree distribution of nodes, $P(k)$, as a function of $k$ for $\zeta = 1000$. We can see that for all values of $\zeta$ we get similar results to those shown in Fig.~3(d) in the main text. We can see that our simulation results (circles) and the analytic solution (continuous lines) overlap for $\zeta = 1000$ and are close enough to lower values of $\zeta$.}
	\label{fig:D2}	
\end{figure}

\section{The effect of different values of $\langle k \rangle$ on the model}
We also tested our model for a different value of average degree, $\langle k \rangle = 3$.
\begin{figure}[h!]
	\centering
	\begin{tikzpicture}[      
	every node/.style={anchor=north east,inner sep=0pt},
	x=1mm, y=1mm,
	]   
	\node (fig1) at (-30,0)
	{\includegraphics[scale=0.11]{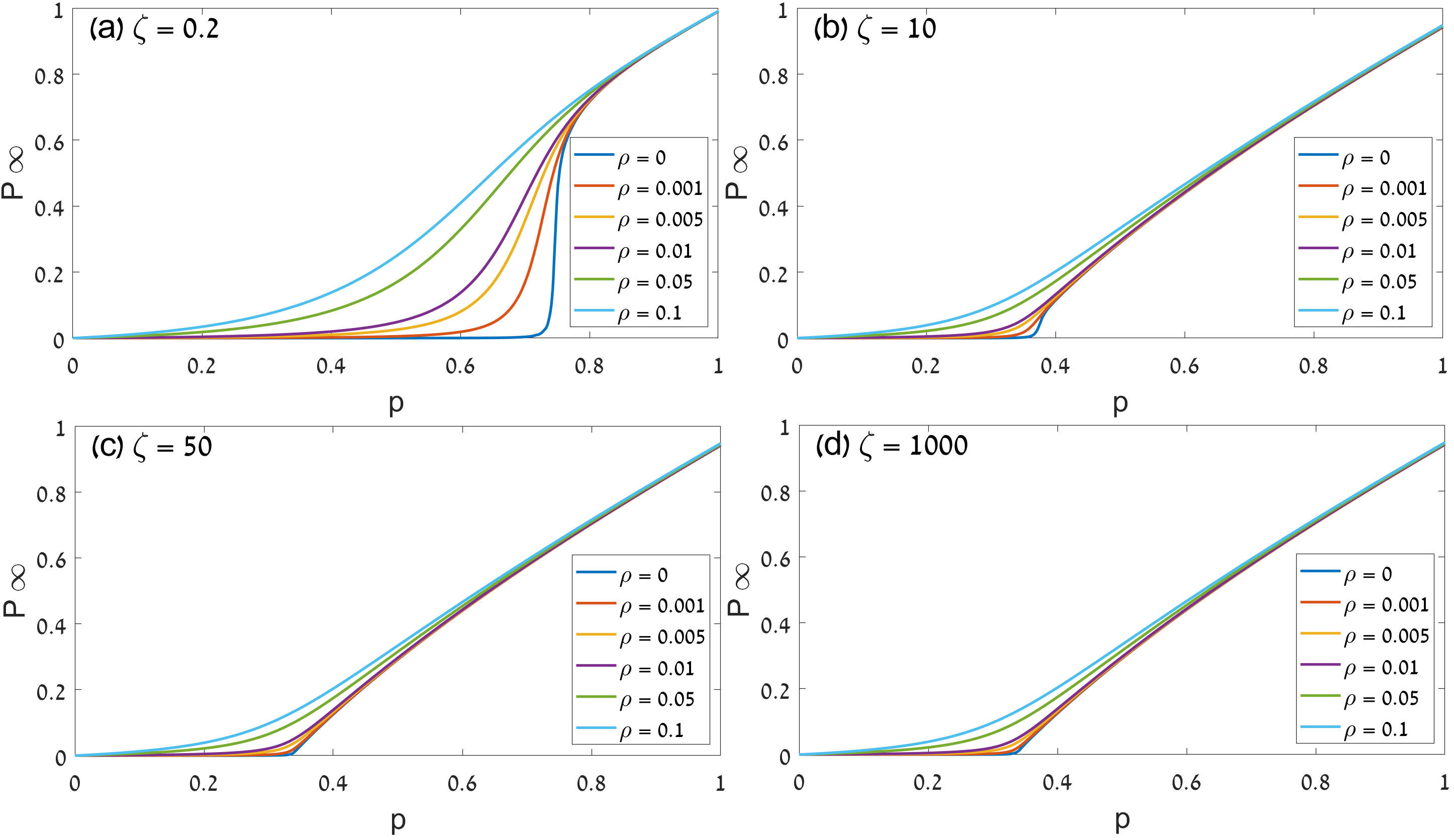}};
	\end{tikzpicture}
	\caption{\textbf{Randomly distributed reinforced nodes in spatial networks with a different average degree, $\mathbf{\langle k \rangle = 3}$.} \textbf{(a)} The giant component, $P_{\infty}$, as a function of $p$ for $\zeta = 0.2$. The phase transition is at $p_c \simeq 0.72$. \textbf{(b)} The giant component, $P_{\infty}$, as a function of $p$ for $\zeta = 10$. \textbf{(c)} The giant component, $P_{\infty}$, as a function of $p$ for $\zeta = 50$. \textbf{(d)} The giant component, $P_{\infty}$, as a function of $p$ for $\zeta = 1000$, here $p_c = 0.33$ (since $\langle k \rangle = 3$). We can see that for all values of $\zeta$ we get similar results to those shown in Fig.~2 in the main text expect for the values of $p_c$. }
	\label{fig:k3gen}	
\end{figure}

In Fig.~D1 we can see that for all values of $\zeta$ we get similar results to the results shown in Fig.~2 in the main text, meaning the effects of randomly distributed reinforced nodes do not change for different values of average degree $\langle k \rangle$. However, there is one exception to these similarities: the values of $p_c$ for each of these values of $\zeta$ are higher, meaning that these spatial networks are less resilient. This is due to the lower values of average degree $\langle k \rangle$, the values of $p_c$ are higher, with the absolute limit of $\langle k \rangle = 1$ where $p_c = 1$ for all values of $\zeta$.

\begin{figure}[h!]
	\centering
	\begin{tikzpicture}[      
	every node/.style={anchor=north east,inner sep=0pt},
	x=1mm, y=1mm,
	]   
	\node (fig1) at (-30,0)
	{\includegraphics[scale=0.11]{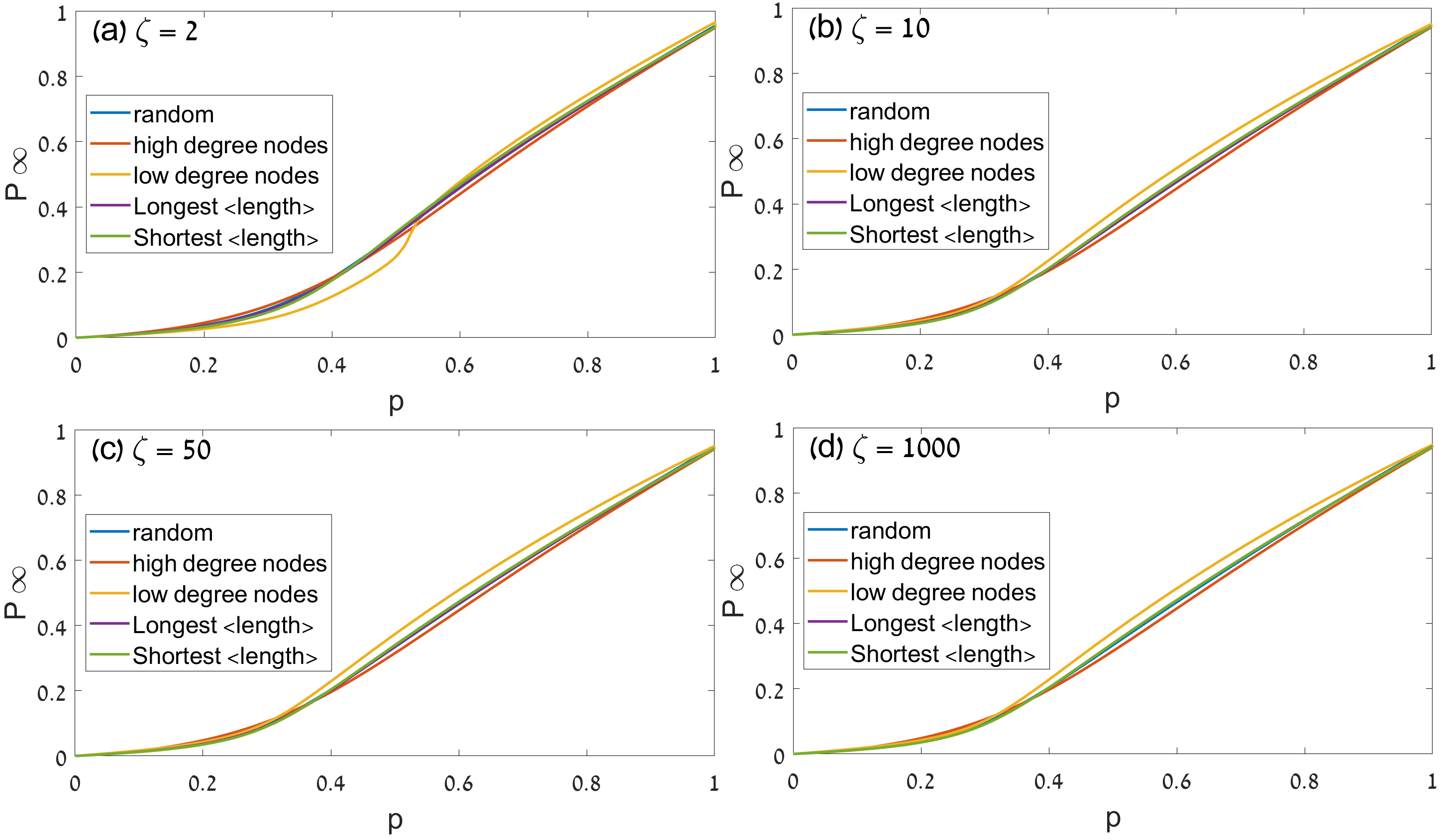}};
	\end{tikzpicture}
	\caption{\textbf{preferred strategies for reinforced distribution in spatial networks with a different average degree, $\mathbf{\langle k \rangle = 3}$.} \textbf{(a)} The giant component, $P_{\infty}$, as a function of $p$ for $\zeta = 2$. \textbf{(b)} The giant component, $P_{\infty}$, as a function of $p$ for $\zeta = 10$. \textbf{(c)} The giant component, $P_{\infty}$, as a function of $p$ for $\zeta = 50$. \textbf{(d)} The giant component, $P_{\infty}$, as a function of $p$ for $\zeta = 1000$. We can see that for all values of $\zeta$ we get similar results to those shown in Fig.~3. The best strategy is again highly dependent on the stages of the percolation. For earlier stages it is best to reinforce low degree nodes, while for later stages it is best to reinforce high degree nodes. }
	\label{fig:k3best}	
\end{figure}

In Fig.~D2 we can see that for all values of $\zeta$ we get similar results to those in Fig.~3 in the main text, meaning the prioritize strategies for reinforced distribution do not change for different values of average degree $\langle k \rangle$ and it still depends on the stages of the percolation process.

\bibliographystyle{unsrt}
\bibliography{mybib}
\end{document}